\documentclass[sigconf]{acmart}

\renewcommand\footnotetextcopyrightpermission[1]{}

\usepackage{graphicx} 
\usepackage{xspace}
\usepackage{enumitem}
\usepackage{tabularx}
\usepackage{makecell}
\usepackage{multirow}
\usepackage{subcaption}

\newcommand{\app}{\textsc{LogiAgent}\xspace}

\newcommand{\ie}{{i.e.,}\xspace}
\newcommand{\eg}{{e.g.,}\xspace}

\newcommand{\scode}[1]{\texttt{#1}\xspace}

\begin{document}
\title[Automated Logical Testing for REST Systems with LLM-Based Multi-Agents]{\app: Automated Logical Testing for REST Systems with \\LLM-Based Multi-Agents}

\author{Ke Zhang}
\affiliation{
\institution{Xidian University}
\country{China}
}
\authornote{Both authors contributed equally to this work.}

\author{Chenxi Zhang}
\affiliation{
\institution{Xidian University}
\country{China}
}
\authornotemark[1]

\author{Chong Wang}
\affiliation{
\institution{Nanyang Technological University}
\country{Singapore}
}
\authornote{Chong Wang is the corresponding author.}

\author{Chi Zhang}
\affiliation{
\institution{Fudan University}
\country{China}
}

\author{YaChen Wu}
\affiliation{
\institution{Fudan University}
\country{China}
}

\author{Zhenchang Xing}
\affiliation{
\institution{CSIRO's Data61}
\country{Australia}
}

\author{Yang Liu}
\affiliation{
\institution{Nanyang Technological University}
\country{Singapore}
}

\author{Qingshan Li}
\affiliation{
\institution{Xidian University}
\country{China}
}

\author{Xin Peng}
\affiliation{
\institution{Fudan University}
\country{China}
}

\begin{abstract}
Automated testing for REST APIs has become essential for ensuring the correctness and reliability of modern web services. While existing approaches primarily focus on detecting server crashes and error codes, they often overlook logical issues that arise due to evolving business logic and domain-specific requirements. To address this limitation, we propose \app, a novel approach for \textbf{logical testing} of REST systems. Built upon a large language model (LLM)-driven multi-agent framework, \app integrates a Test Scenario Generator, API Request Executor, and API Response Validator to collaboratively generate, execute, and validate API test scenarios. Unlike traditional testing methods that focus on status codes like \scode{5xx}, \app incorporates \textbf{logical oracles} that assess responses based on business logic, ensuring more comprehensive testing. The system is further enhanced by an Execution Memory component that stores historical API execution data for contextual consistency. We conduct extensive experiments across 12 real-world REST systems, demonstrating that \app effectively identifies 234 logical issues with an accuracy of 66.19\%. Additionally, it basically excels in detecting server crashes and achieves superior test coverage compared to four state-of-the-art REST API testing tools. An ablation study confirms the significant contribution of \app’s memory components to improving test coverage.
\end{abstract}

\maketitle

\section{Introduction}
REST (Representational State Transfer) has become the dominant architectural style for web services, providing a scalable and flexible approach to system integration through stateless interactions and standardized HTTP methods~\cite{fielding2002principled}. REST APIs enable interoperability across distributed systems, forming the backbone of modern cloud computing platforms, microservices architectures, and the Web of Things~\cite{cloud2011nist,newman2021building,guinard2010resource}. However, ensuring the correctness, robustness, and reliability of REST systems remains a significant challenge due to evolving API specifications, complex dependencies, and the potential for inconsistent behaviors across different execution environments~\cite{TSC_rest,rest_pattern,tse_micro,API_evol}. 

Automated REST API testing has become a vital solution to address these challenges, with recent research investigating methods that leverage large language models (LLMs) to improve test case generation~\cite{RestGPT, AutoRestTest,RESTSpecIT}. 
RestGPT~\cite{RestGPT} and AutoRestTest~\cite{AutoRestTest} leverage LLMs to identify constraints and generate valid parameter values.
RESTSpecIT~\cite{RESTSpecIT} utilizes LLMs to analyze API responses to automatically infer API specification in addition to parameter generation.
These works alleviate the problem of traditional methods that frequently generate invalid parameters.
Despite the advances, they struggle to accurately capture real-world business logic, focusing on detecting \scode{5xx} errors while overlooking logical issues.

To this end, we propose conducting \textbf{logical testing} for REST systems to expand the scope of existing research on automated REST API testing. Unlike traditional testing objectives (e.g., \textit{fuzz testing} for server crashes), logical testing seeks to address two key limitations. First, rather than directing test generation towards deeper exploration based on \scode{404} or \scode{5xx} feedback, logical testing focuses the exploration on more complex, realistic scenarios guided by business logic feedback. Second, the validation of API request responses should not only consider status codes like \scode{5xx}, but also include \textbf{logical oracles} that reflect the business logic and domain knowledge. For example, some API requests may return responses with a \scode{200 OK} status, but the information in the response body may be logically incorrect given the specific scenario context.

In this paper, we introduce \app, a novel automated logical testing approach for REST systems based on an LLM-driven multi-agent system. The framework consists of three core agents: {Test Scenario Generator}, {API Request Executor}, and {API Response Validator}, which collaboratively generate, execute, and validate API test scenarios. A global {Scenario Scheduler} manages the execution status, while a long-term {Execution Memory} stores historical API execution data to maintain synchronization and contextual consistency. The process begins with the Test Scenario Generator, which creates a test scenario and adds it to the Scenario Scheduler. The workflow then cycles between the API Request Executor, which retrieves and executes test steps, and the API Response Validator, which validates the responses. Based on the validation results, the Scenario Scheduler updates the status and checks for scenario termination. Once terminated, the Test Scenario Generator creates a new scenario, and the process repeats.

We conduct extensive experiments to evaluate the effectiveness of \app in detecting both logical errors and server crashes across 12 real-world REST systems. The results show that \app effectively detects logical issues, identifying 234 issues (139 bugs and 95 enhancements) with a reasonable accuracy of 66.19\%. Additionally, \app demonstrates leading performance in detecting \scode{500}-code server crashes (49 errors identified in total) and achieves superior test coverage, compared to four state-of-the-art REST API testing tools. Finally, the results of an ablation study confirm the contribution of the memory components in \app to improved test coverage.

To summarize, this paper makes the following key contributions:
\begin{itemize}[leftmargin=12pt, topsep=0.5em]
    \item \app, a novel automated logical testing approach for REST systems based on an LLM-driven multi-agent system, which extends the scope of existing research on automated REST API testing. The system includes three LLM-based agents, a global Scenario Scheduler, and long-term Execution Memory to collaboratively execute the testing workflow.
    \item Experimental results demonstrating the effectiveness of \app in detecting both logical issues and server crashes in REST systems. It identifies 234 logical issues and 49 server crashes across 12 real-world REST systems.
\end{itemize}
\section{Preliminaries and Motivation}
In this section, we first introduce the fundamentals of logical testing for REST systems and contrast it with traditional REST API fuzz testing. We then present motivating examples to highlight the challenges involved in logical testing for REST systems.

\subsection{Logical Testing for REST Systems}

\textbf{Business Scenario Orchestration.}
In logical testing, the goal is to model real-world workflows and use cases that the APIs are designed to support, simulating practical business scenarios to ensure that APIs are called as intended and validated under real-world conditions. While traditional fuzz testing also generates API call chains, its primary focus is on creating diverse API request parameters and triggering various and deep execution paths, without guaranteeing that these paths align with the way real users interact with the system. Moreover, fuzz testing typically sends random or edge-case data to identify problems like server crashes or failures (e.g., \scode{5xx} error codes), whereas logical testing aims to uncover logical issues that extend beyond server crashes. For example, in an e-commerce platform, a business scenario might involve steps such as creating a user profile, adding products to a shopping cart, placing an order, and processing payment. Logical testing ensures that these steps are executed as orchestrated, verifying that the API functions consistently and as intended across multiple interconnected endpoints, while adhering to real-world business logic. This requires a higher level of orchestration of REST APIs to uncover more realistic issues that would be triggered by general users.

\textbf{Oracle and Response Validation.}  
In logical testing, an oracle defines the expected behavior or output of an API in response to a given input. Unlike traditional fuzz testing, which typically focuses on basic checks (\eg validating if a response code falls within a specific range like \scode{5xx}), logical testing requires more advanced oracles. These oracles go beyond checking the response code; they assess the content, structure, and relationships within the response to ensure they align with the intended business logic. The oracle is derived not only from the API's technical specifications but also from a deep understanding of the semantic and logical relationships embedded in business rules, domain knowledge, and use-case scenarios. Effective response validation, based on these oracles, ensures that the API functions as intended under various test conditions, identifying logical discrepancies that might otherwise be overlooked in traditional crash-focused fuzz testing.
For example, when creating a user profile through an API, the oracle defines the expected data format (\eg name, age, address) and ensures that the response aligns with the semantic meaning of the data. This involves checking that the returned data accurately reflects the input and adheres to business rules—such as ensuring the user's age is logically consistent with other fields (\eg age should not be negative or unrealistic) and that the user's address conforms to valid real-world standards (\eg proper address format). 

\subsection{A Motivating Example}
We use the PetStore system~\cite{petstore} as an example to illustrate the motivation behind logical testing and highlight key challenges. Consider a multi-step business scenario: a user registers, logs in, creates a pet profile, deletes the pet, and then attempts to place an order for the deleted pet. Executing these API calls in PetStore reveals three logical issues. 

\begin{figure}
    \centering
    \begin{subfigure}{0.95\columnwidth}
        \centering
        \includegraphics[width=\linewidth]{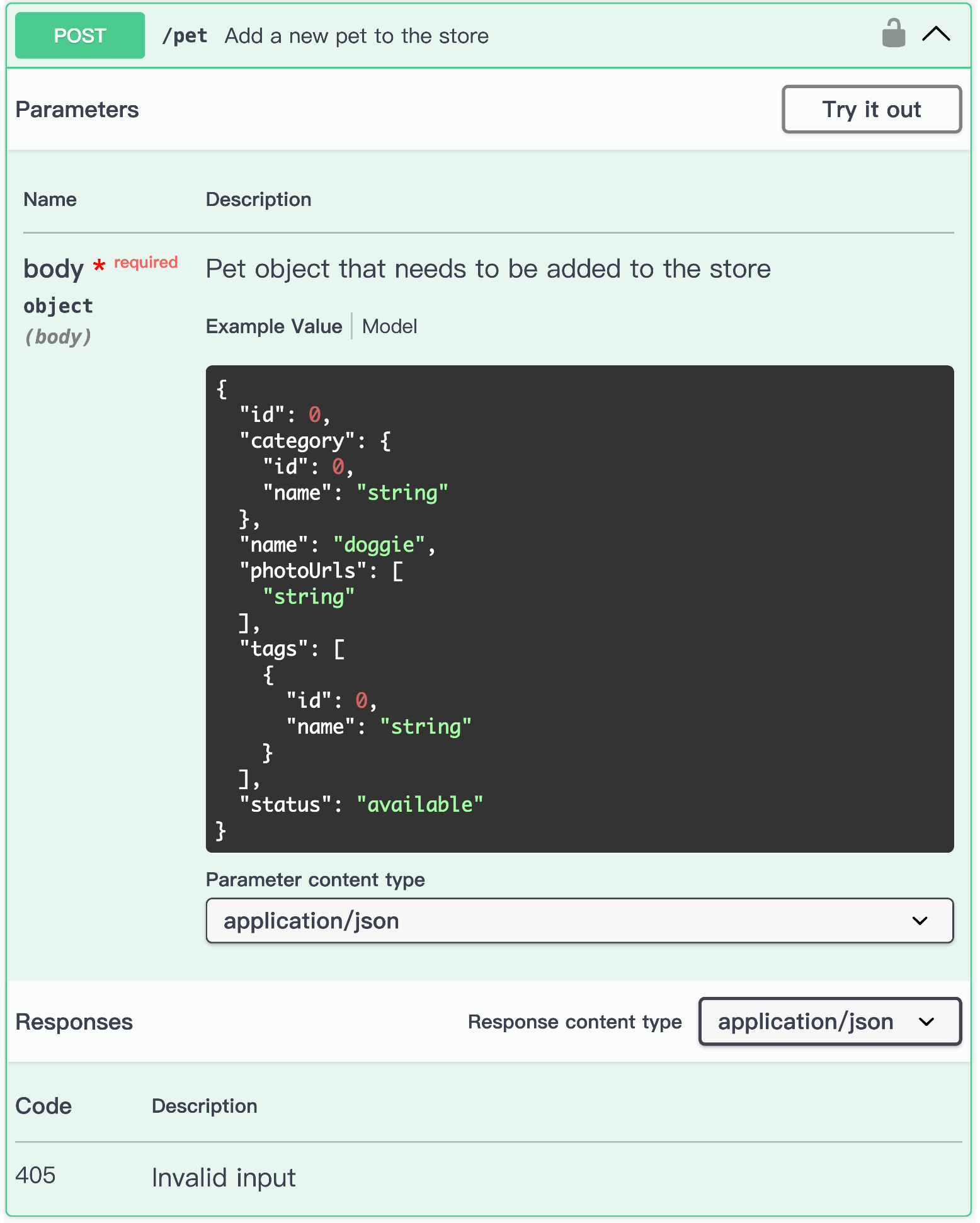}
        \caption{The OpenAPI Specification}
        \label{fig:motivation_pet}
    \end{subfigure}
    \begin{subfigure}{0.95\columnwidth}
        \centering
        \includegraphics[width=\linewidth]{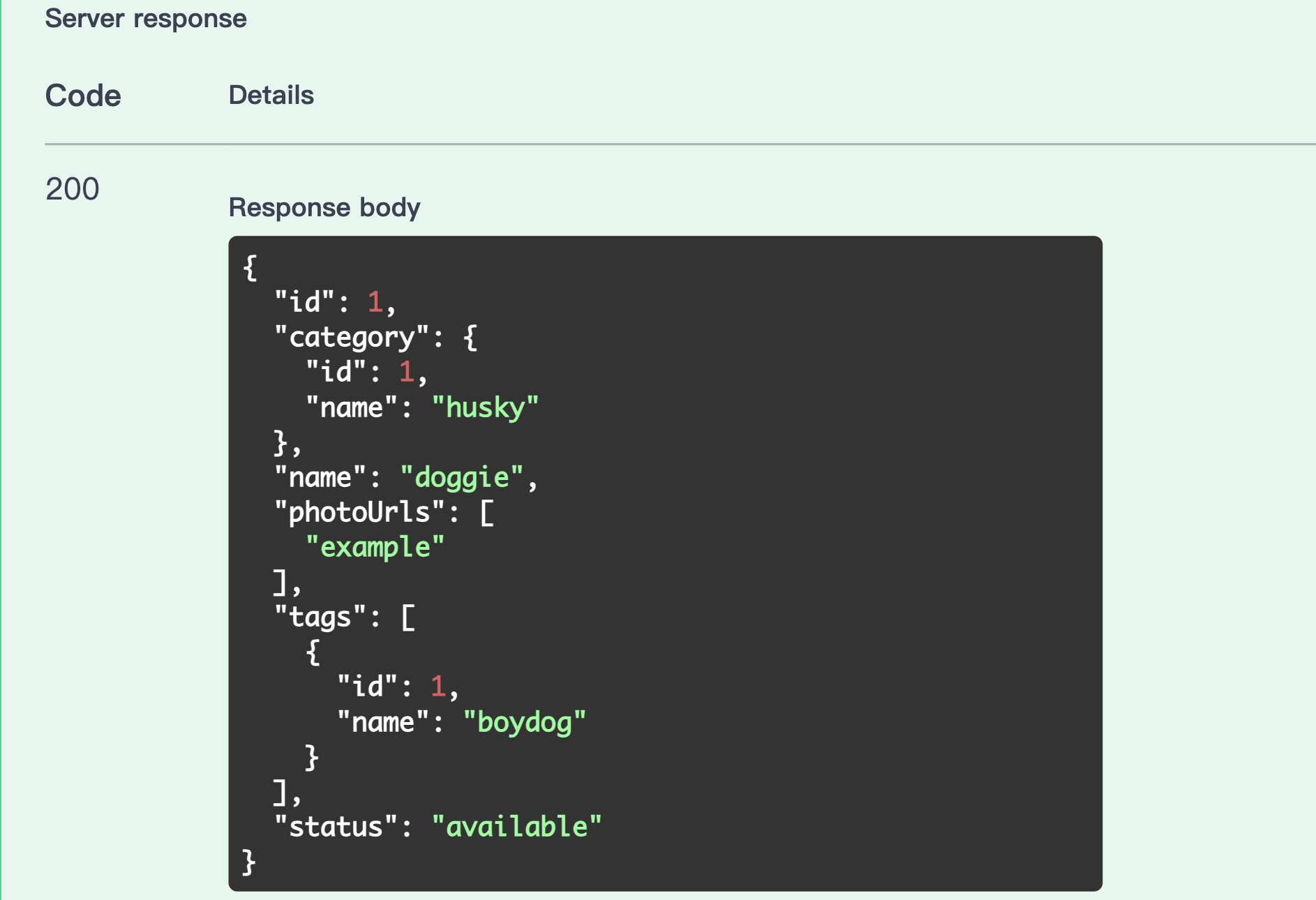}
        \caption{An Example Response}
        \label{fig:motivation_pet_resp}
    \end{subfigure}
    \caption{The OpenAPI Specification and Example Response of \texttt{/pet} API}
    \label{fig:motivation_pet_all}
    \vspace{-5mm}
\end{figure}

First, when creating a pet profile, the ``\texttt{POST /pet}'' API (Figure \ref{fig:motivation_pet}) is used to add a new pet with details such as category, name, photo URLs, and tags. However, sending a request with an invalid photo URL (\eg ``example'') results in a \scode{200 OK} response (Figure \ref{fig:motivation_pet_resp}), indicating successful execution. This exposes two issues: (1) By REST design conventions, a successful creation via \texttt{POST} should return \scode{201 Created} rather than \scode{200 OK}~\cite{RFC9110}; (2) The system incorrectly accepts invalid photo URL, leading to potential data integrity and accessibility problems. 
A more severe logical issue arises in the final step when creating an order for a deleted pet. Despite the pet no longer existing, the system incorrectly returns a \scode{200 OK} response, causing a logical inconsistency that may lead to critical vulnerabilities. 

This example highlights three key insights. First, logical testing is essential for identifying both minor inconsistencies and severe business logic flaws in REST systems. Second, certain logical flaws only emerge through carefully constructed sequences of operations, underscoring the need for complex and coherent test scenarios. Third, validating API responses requires robust oracles derived from multiple sources beyond API documentation, including business logic, domain knowledge, and best practices.

\section{Approach}
In this paper, we propose \app, an approach to automated logical testing for REST systems using a multi-agent system powered by large language models.

\subsection{Overview}
\begin{figure*}
    \centering
    \includegraphics[width=0.9\textwidth]{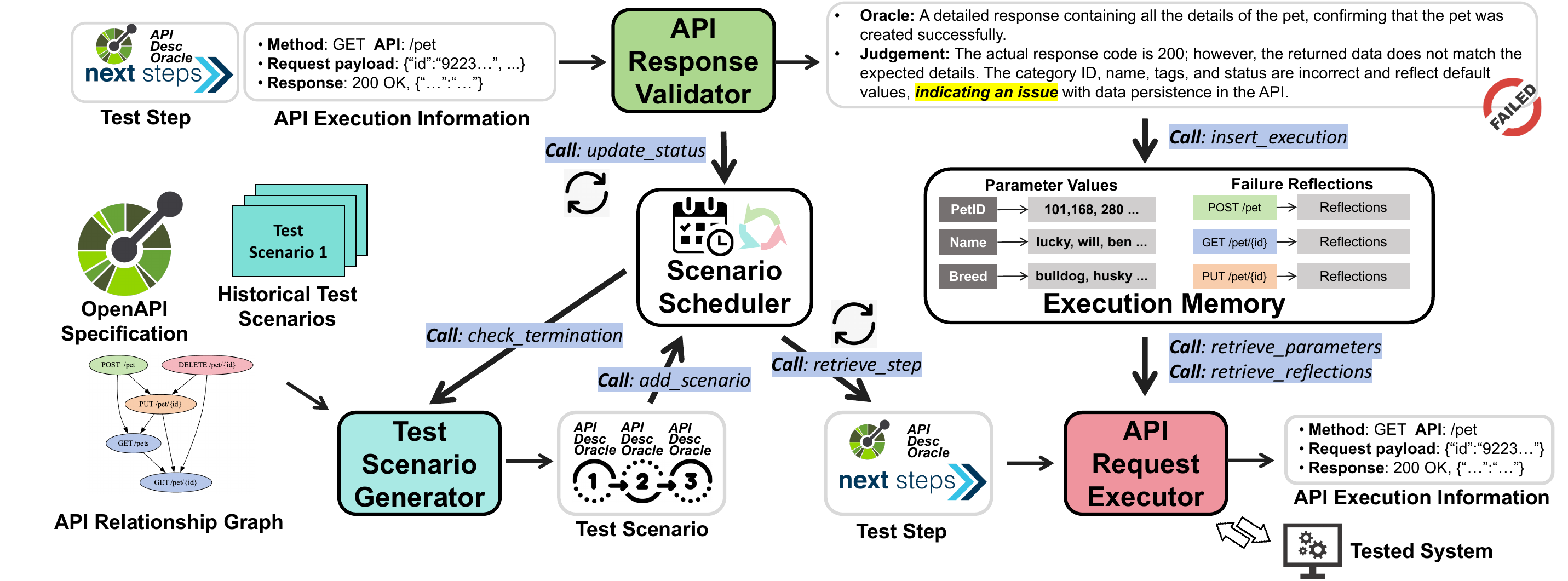}
    \vspace{-2mm}
    \caption{Approach Overview of \app}
    \label{fig:agent-main}
    \vspace{-2mm}
\end{figure*}

As shown in Figure \ref{fig:agent-main}, the system comprises three primary LLM-based agents: \textbf{Test Scenario Generator}, \textbf{API Request Executor}, and \textbf{API Response Validator}. These agents collaboratively generate, execute, and validate API test scenarios. Additionally, a global \textbf{Scenario Scheduler} maintains the execution status of the current test scenario, while a long-term \textbf{Execution Memory} stores historical API request executions, ensuring synchronized scheduling and contextual consistency across agents. In this section, we first introduce the key components of the multi-agent system before detailing the overall workflow. The details of the three agents are introduced in the following sections.

\subsubsection{LLM-based Agents}
The following describes the roles and functionalities of the three LLM-based agents:
\begin{itemize}[leftmargin=12pt]  
    \item \textit{Test Scenario Generator} leverages an LLM to create comprehensive and diverse test scenarios that accurately reflect the business logic of the target REST system. Each time it runs, the generator prompts an LLM to produce a new test scenario consisting of a sequence of \textbf{Test Steps}, each paired with its expected behavior as defined by \textbf{Test Oracles}. A test step includes an API to execute and a brief description, while an oracle is derived from multiple sources, including API documentation (\eg expected response descriptions), the scenario context (\eg business logic), and general LLM knowledge (\eg best practices beyond the API documentation). To ensure that generated scenarios align with real-world business logic, the generator utilizes the \textbf{API Relationship Graph (ARG)}, an undirected graph constructed from API documentation that encodes semantic relationships between APIs. Additionally, to promote diversity and minimize redundancy, it considers previously generated test scenarios and their execution results stored in the {Execution Memory}.

    \item \textit{API Request Executor} is responsible for constructing request payloads (API parameters) for a test step using an LLM and executing the API request. To maintain contextual consistency (\eg ensuring the same user ID across API requests), the executor considers previous API requests and responses within the same scenario (\ie a short memory specific to the current scenario) before prompting the LLM. Additionally, it retrieves relevant historical execution information from the long-term {Execution Memory}, including reference parameter values and prior failure reflections. This helps the agent avoid known valid or invalid parameter values when constructing requests. Once the API request is prepared, the executor invokes the \textit{do\_request} function to execute it and collects the response.

    \item \textit{API Response Validator} employs an LLM to assess whether the collected API response aligns with the expected behavior (\ie oracle) based on the current scenario context (\eg business logic and execution state) and the LLM’s text understanding capabilities. If the response violates the oracle, an issue—ranging from server crashes to logical inconsistencies—is reported. The API request details and failure reflections (\ie explanations generated by the LLM) are also recorded in the {Execution Memory} to guide future request optimizations.
\end{itemize}  

\subsubsection{Scenario Scheduler}
When testing a target REST system, a key challenge is managing the execution status of scenarios and steps while coordinating the execution flow among agents. Specifically, the multi-agent system must determine when a scenario has been fully executed or has failed, requiring the generation of a new scenario. It also needs to assess whether a specific step has passed or requires re-execution. To address this, we design a global {Scenario Scheduler} to enhance status awareness within the multi-agent system. The scheduler maintains a list of test steps and a pointer to the current step being executed. Each step has a \textit{retry} property indicating the number of re-execution attempts.

\begin{itemize}[leftmargin=12pt]
    \item \textit{Scenario Addition} (\textit{add\_scenario}): Each time the {Test Scenario Generator} creates a new test scenario, its steps are sequentially added to the scheduler. The pointer is set to the first step, and each step's \textit{retry} count is initialized to zero.
    \item \textit{Step Retrieval} (\textit{retrieve\_step}): The scheduler returns the step currently pointed to. Notably, the retrieved step may be the same as in the previous retrieval if a failure occurred, necessitating a step re-execution.
    \item \textit{Status Update} (\textit{update\_status}): After validating the current API request, the scheduler updates the execution status. If the validation confirms that the step has passed, it is removed from the step list, and the pointer advances to the next step. Otherwise, the \textit{retry} count is incremented. If \textit{retry} exceeds a predefined limit (\eg 3), the step list is cleared, and the scenario is marked as a failed execution.
    \item \textit{Termination Check} (\textit{check\_termination}): Determines whether the current scenario should be terminated and a new one generated. Termination is triggered when the step list becomes empty.
\end{itemize}

\subsubsection{Execution Memory}
To ensure sufficient exploration depth and test coverage, it is crucial for the {API Request Executor} to generate valid and diverse parameter values. However, LLMs struggle to produce such inputs effectively without execution feedback. To address this, we introduce an {Execution Memory} to facilitate information sharing between the {API Request Executor} and {API Response Validator}. Specifically, this memory maintains records of:
(i) parameter names and values (\eg \textit{product\_status}=\textit{`sold'}) for requests whose responses successfully pass validation, and
(ii) failure reflections for requests whose responses fail validation, where the reflections are generated by LLMs in the {API Response Validator}.

\begin{itemize}[leftmargin=12pt]
    \item \textit{Feedback Insertion} (\textit{insert\_execution}): Each time the {API Response Validator} completes validation for an API response, execution feedback is inserted into memory. If the validation is passed, the parameter names and values from the API request are stored as key-value pairs. Otherwise, a failure reflection record is inserted, where the API serves as the key, and the corresponding API request details and failure explanation are stored as the value.
    \item \textit{Parameters Retrieval} (\textit{retrieve\_parameters}): Given an API, the BM25 (Best Matching 25) retrieval algorithm~\cite{BM25} filters relevant records from memory to provide reference values for its parameters. The retrieval query combines the API specification's description text with the current step description, retrieving the top 10 most relevant previously stored parameter names and values. Notably, the retrieved parameters are not necessarily from the current API being executed, as semantically similar parameter names across different APIs enable a degree of parameter value generalization.
    \item \textit{Reflections Retrieval} (\textit{retrieve\_reflections}): Given an API, this function returns stored failure reflection records of this API. These reflections help accumulate system knowledge over time, reducing the likelihood of repeating the same errors and improving the effectiveness of future executions.
\end{itemize}

\subsubsection{Workflow}
As illustrated in Figure~\ref{fig:agent-main}, given a target REST system, the process begins with the Test Scenario Generator, which generates a test scenario and adds it to the {Scenario Scheduler} by calling the \textit{add\_scenario} function.  
Next, the workflow enters a cycle between the {API Request Executor} and the {API Response Validator} to execute and validate the scenario. Specifically, the {API Request Executor} continuously retrieves the current step to execute from the {Scenario Scheduler} by calling the \textit{retrieve\_step} function, executes the step, and collects the response. The collected response is then passed to the {API Response Validator} for validation. Based on the validation results, the {Scenario Scheduler} updates the execution status by calling the \textit{update\_status} function and checks whether the current scenario is terminated using the \textit{check\_termination} function.  
This cycle repeats until the \textit{check\_termination} function returns true, at which point the {Test Scenario Generator} is reactivated to generate a new test scenario, and the {Scenario Scheduler} begins managing the newly generated scenario.


\subsection{Test Scenario Generator}
The Test Scenario Generator creates test scenarios by leveraging an LLM to analyze API documentation and apply its acquired domain knowledge, allowing for a deeper understanding of the business logic behind the target REST system.

\subsubsection{API Relationship Graph}
While the API documentation (OpenAPI Specification Format~\footnote{https://swagger.io/specification/}) provides detailed descriptions of individual APIs, its raw organizational structure makes it challenging for LLMs to capture the semantic relationships between them. The documentation typically presents APIs as isolated entities, each with a brief functionality description, parameter details (with examples), and response information. However, real-world business scenarios often involve sequences of API calls that encode underlying semantic logic. To bridge this gap, we introduce the \textbf{API Relationship Graph (ARG)}, an undirected graph that models API interactions and dependencies, enhancing the global understanding of APIs within the target REST system.

To construct the ARG, we employ a two-phase process to identify semantically related API pairs. In the first phase, we iterate through all possible API pairs and compute their semantic similarity. Each API is encoded into a vector representation using the OpenAI embedding model~\cite{openai_embedding}, which processes the API's description text. The semantic similarity between two APIs is then measured using cosine similarity, and only pairs that exceed a predefined threshold (\eg 0.5) are retained.  
In the second phase, for each retained API pair, an LLM (\eg GPT-4o-mini) analyzes their API documentation to determine whether they exhibit logical interactions or dependencies. Through this process, the ARG is constructed by representing APIs as nodes and inserting edges between related API pairs. Note that if isolated APIs exist, we connect them to all other APIs by adding edges. Each API node in the ARG is further enriched with attributes such as request method (\eg \texttt{GET} and \texttt{POST}), path, and a structured Markdown-formatted description.  
The first phase serves as a necessary filtering step, as when the number of APIs grows, making direct pairwise evaluation with LLMs impractical due to computational constraints. 

Based on the constructed ARG, we apply \textbf{random walk} (maximum length is 10) to dynamically select a subset of interconnected APIs that capture semantic aspects of the target REST system. Compared to directly feeding all APIs into LLMs for test scenario generation, this approach exposes LLMs to diverse API combinations while ensuring focus on relevant APIs and reducing context overload.

\subsubsection{Agent Initialization}
\begin{figure}
    \centering
    \includegraphics[width=0.9\linewidth]{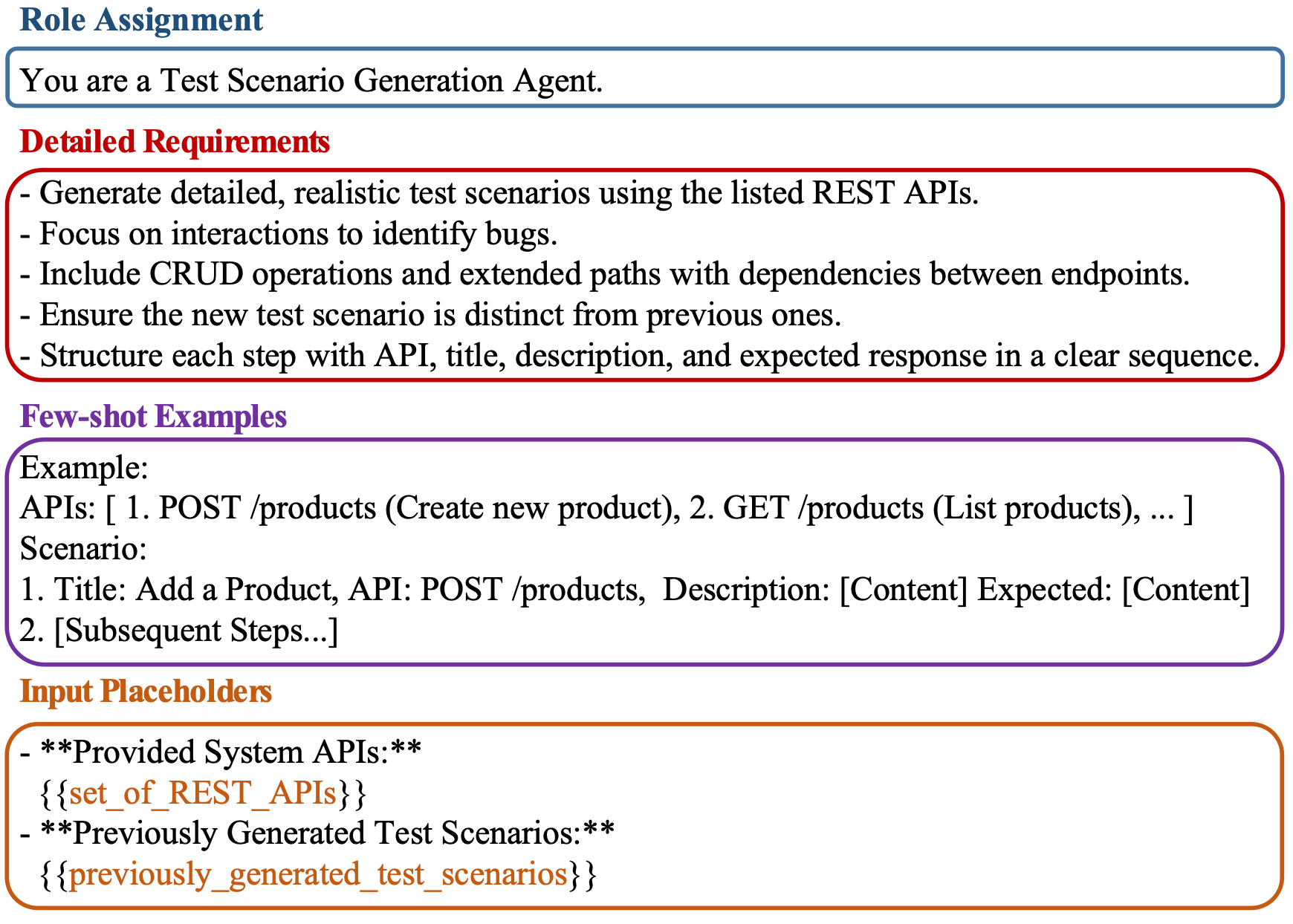}
    \caption{Prompt Template Used in Test Scenario Generator}
    \label{fig:test_scenario_prompt}
    \vspace{-5mm}
\end{figure}

The prompt template shown in Figure \ref{fig:test_scenario_prompt} is designed to initialize the LLMs as high-quality test scenario generators. Note that the prompt has been simplified for clarity, while the complete version is available in our replication package~\cite{replication}.

Specifically, the prompt template comprises four key components:  
\textbf{Role Assignment}, which defines the LLM’s persona as a specialized agent for generating test scenarios;  
\textbf{Detailed Requirements}, which outlines specific criteria to ensure the generation of high-quality test scenarios. For a REST system, an effective test scenario should not only simulate real-world business logic but also introduce new situations that have not yet been tested. 
To achieve this, we encode these requirements into the instructions and specify a structured output format where each step includes a Title, API, Description, and Expected Response (Oracle);
\textbf{Few-shot Examples}, which provide carefully crafted example test scenarios inspired by in-context learning~\cite{icl_rethinking,icl_survey}, helping the LLM understand the task requirements and expected format; and
\textbf{Input Placeholders}, which supplies a set of APIs selected through random walk on the ARG, along with previously generated test scenarios, to prevent redundancy and encourage diversity in scenario generation.

\subsubsection{Test Scenario Generation}
Each time the agent executes, the input placeholders in the prompt template are filled with the required inputs. First, a set of relevant APIs is collected by applying random walk on the ARG and inserted into the corresponding placeholder. Then, historical test scenarios are fed into the corresponding placeholder. To ensure the context window size is within limits, no more than ten historical scenarios are provided, selected randomly. Once the inputs are filled into the prompt template, the agent requests the LLMs to generate a structured test scenario, typically consisting of 8 to 12 test steps. Each step includes the relevant API information, a brief description of the step, and the expected response of the request (\ie oracle). The generated scenario is subsequently sent to the Scenario Scheduler.

\subsection{API Request Executor}
The API Execution Agent is the agent responsible for generating API request payload and executing actual API request.

\begin{figure}
    \centering
    \includegraphics[width=0.9\linewidth]{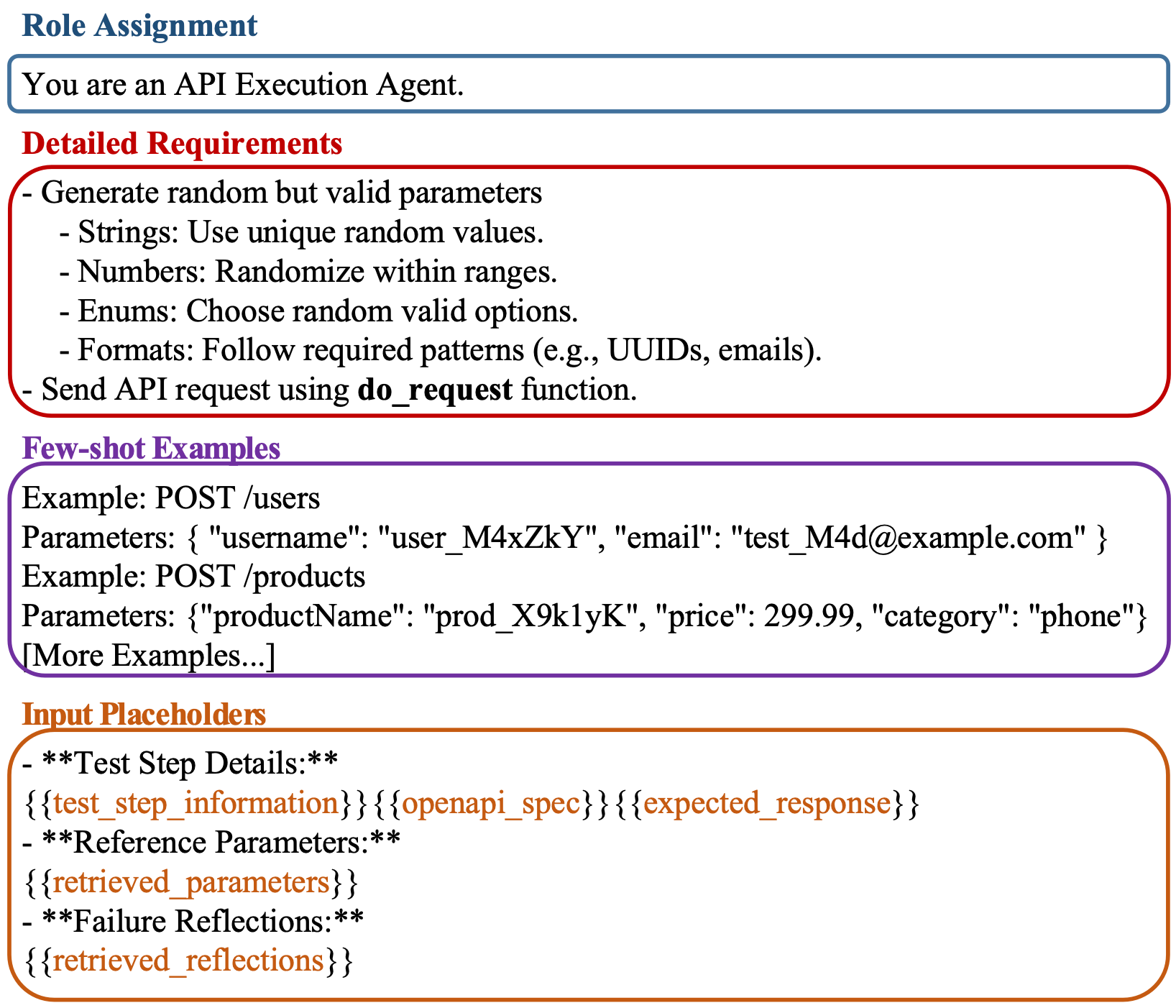}
    \caption{Prompt Template Used in API Request Executor}
    \label{fig:api_execution_prompt}
    \vspace{-5mm}
\end{figure}

\subsubsection{Agent Initialization}
This executor agent is initialized using a four-component prompt template, as shown in Figure \ref{fig:api_execution_prompt}.  
\textbf{Role Assignment} defines the agent's primary function as an API execution handler; 
\textbf{Detailed Requirements} specify the API request execution using \textit{do\_request} function and parameter generation rules for various data types with their validation patterns;
\textbf{Few-shot Examples} demonstrate well-designed API request parameters examples, helping the LLM understand how to generate random and valid parameters for APIs; and
\textbf{Input Placeholders} supply the necessary details for executing each step.

\subsubsection{Test Step Retrieval}
The agent first utilizes the \textit{retrieve\_step} function tool provided by the {Scenario Scheduler} to retrieve the next step for execution, which may be the same as in the previous iteration if a verification failure occurred.

The retrieved test step consists of three primary components, with historical information gathered from {Execution Memory}:
\begin{itemize}[leftmargin=12pt]  
    \item \textit{Test Step Details}: This section includes the test step name, API, test description, API specifications in Markdown format, and the expected response. It serves as the fundamental context for executing the current test step. Notably, the expected response (\ie the oracle) is also provided as auxiliary information to help the LLM understand the step's intent.
    \item \textit{Reference Parameters}: To generate parameter values efficiently and effectively, relevant historical parameter values are retrieved from the Execution Memory by calling its \textit{retrieve\_parameters} function as references.
    \item \textit{Failure Reflections}: Similarly, relevant failure reflection information is retrieved from the {Execution Memory} by calling the \textit{retrieve\_reflections} function to guide the LLM in generating valid parameters, improving the likelihood of a successful API request and enhancing exploration depth.
\end{itemize}
The retrieved information is then filled into {Input Placeholders} to construct a concrete prompt.

\subsubsection{Request Execution}
After generating the parameters, the agent constructs the API payload and sends the request to the target REST system using the \textit{do\_request} function call, then collects the response. 
The \textit{do\_request} function is implemented as a Python tool that provides a clean interface for the agent.

\subsection{API Response Validator}
After each request execution, the API Response Validator evaluates whether the response aligns with expectations and provides an explanation for its judgment.

\begin{figure}
    \centering
    \includegraphics[width=0.9\linewidth]{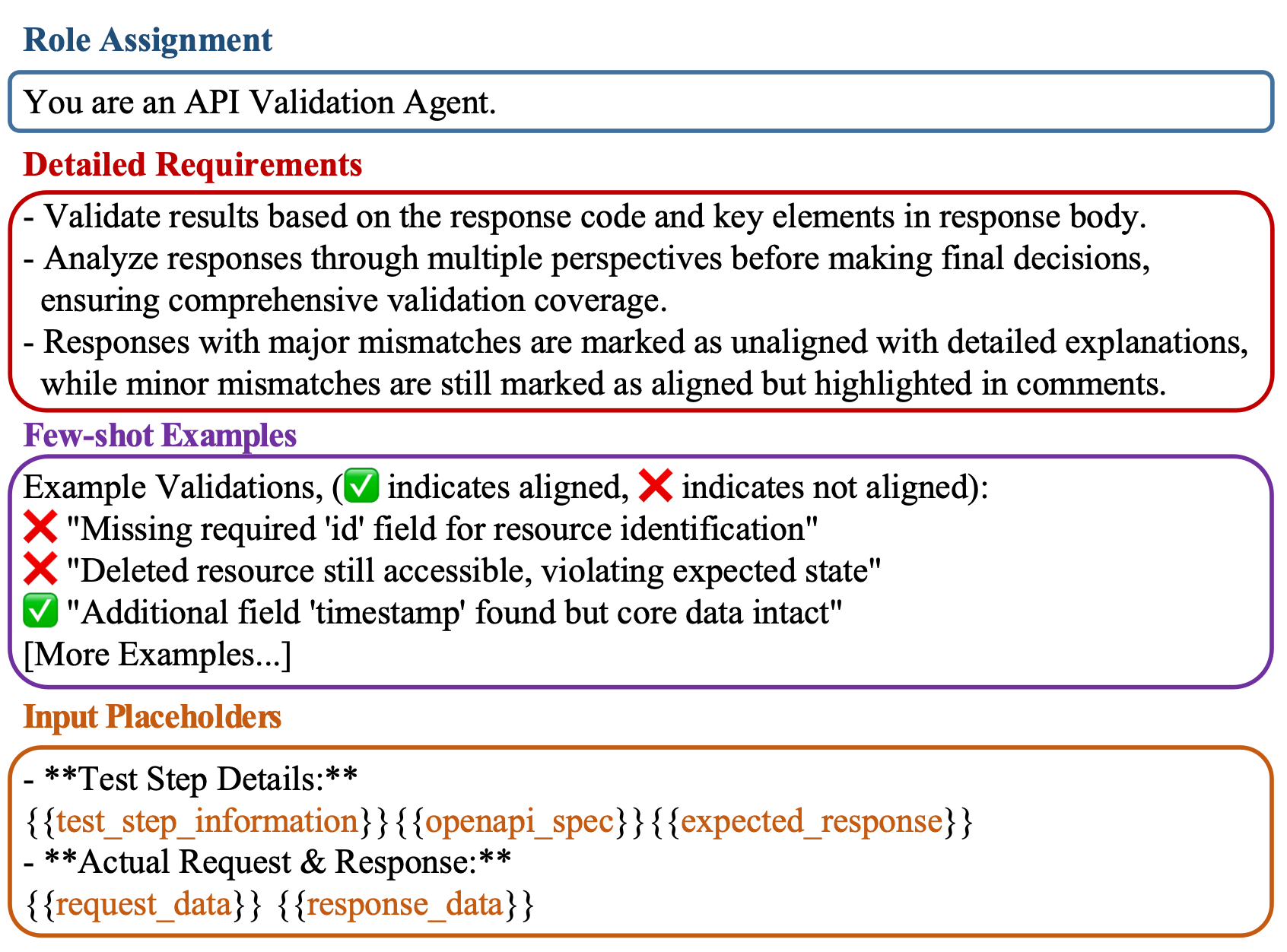}
    \caption{Prompt Template Used in API Response Validator}
    \label{fig:api_validation_prompt}
    \vspace{-5mm}
\end{figure}

\subsubsection{Agent Initialization}
As shown in the figure \ref{fig:api_validation_prompt},  the agent’s prompt template consists of four components: 

\textbf{Role Assignment} defines the agent's primary function as an API Response Validator;
\textbf{Detailed Requirements} specify the validation process through response analysis from multiple perspectives and define the validation criteria for response assessment;
\textbf{Few-shot Examples} demonstrate typical validation scenarios with clear alignment indicators, helping the LLM understand how to assess different types of response mismatches; and
\textbf{Input Placeholders} supply the necessary context including test step details and actual request-response data for validation.


\subsubsection{Validation Process}
In the validation process, the agent first fills required information into \textbf{Input Placeholders} in the prompt template.
To ensure clarity, this not only provides the API response code and response body, but also provides the current API information in Markdown format, along with the request payload and parameters  during request execution. 
This guarantees that the evaluation is based on complete and structured context, facilitating accurate validation.

After gathering all necessary context, the API Validation Agent employs a multi-perspective analysis approach when validating responses.
By evaluating multiple aspects of the response simultaneously, the agent ensures comprehensive validation coverage.
This approach leads to more thorough validation results, where responses with major mismatches are marked as unaligned with detailed explanations, while minor mismatches are documented in the reasoning process.

If the agent detects any issues, such as server crashes or logical inconsistencies, it reports them. The \textit{Execution Memory} is then updated based on the validation results by invoking the \textit{insert\_execution} function.


\section{Evaluation}

We conduct extensive experiments to evaluate the effectiveness of the proposed \app in testing real-world REST systems.

First, we run \app and 4 state-of-the-art baselines on 12 real-world REST systems and record the reported problems, including both \textit{logical issues} and \textit{server crashes}. We assess \app's precision in detecting logical issues through manual annotation (\textbf{RQ1}) and evaluate its effectiveness in detecting server crashes by comparing it against the baselines (\textbf{RQ2}). Second, we analyze the test coverage achieved by \app on the tested REST systems and compare it with the baselines (\textbf{RQ3}). Third, we perform an ablation study to investigate the contribution of the memory components in \app (\textbf{RQ4}). 

All research questions are listed as follows:
\begin{itemize}[leftmargin=15pt]
    \item \textbf{RQ1 (Effectiveness in Detecting Logical Issues)}: How effectively does \app identify logical issues?  
    \item \textbf{RQ2 (Comparison in Detecting Server Crashes)}: How does \app perform in detecting server crashes (\scode{500} Internal Server Error) compared to the baselines?  
    \item \textbf{RQ3 (Test Coverage)}: How does \app's test coverage, including operation coverage and code coverage, compare to the baselines?  
    \item \textbf{RQ4 (Ablation Study)}: What is the impact of \app's memory components on its performance?
\end{itemize}

\subsection{Experimental Setup}

\subsubsection{Tested Systems}

For evaluating \app, we select a diverse set of 12 REST systems, including both online and local deployments, to ensure a comprehensive assessment of our approach.
As shown in Table \ref{table:services}, these systems vary significantly in size, ranging from 556 to 677,521 lines of code, and complexity, containing between 2 and 67 API operations.

Three widely recognized online systems are included to benchmark performance against established systems: Petstore~\cite{petstore}, the official Bills API of the UK Parliament~\cite{billservice}, and the publicly accessible Genome Nexus~\cite{genomenexus}.

The remaining nine systems—Features-Service, REST Countries, News Service, SCS, NCS, Language Tool, Person-Controller, Project-Tracking-System, and User Management—are selected from a recent study~\cite{restgo} and deployed in our local environment. Note that we exclude ten REST systems tested in~\cite{restgo} from our evaluation due to various technical constraints. Four systems (Erc20 Rest Service, Spring Boot Actuator, Proxyprint, and Scout API) fail to provide reliable responses—two due to outdated dependencies incompatible with current external systems, and the other two because of authentication challenges that prevent meaningful testing. CatWatch and CWA Verification are omitted due to stringent rate-limiting mechanisms that significantly hinder the testing process and prevent timely result collection. Problem Controller and Spring Batch REST are excluded as their effective testing requires specialized domain expertise, which would introduce variables unrelated to our tool’s performance. Finally, OCVN and Market Service present substantial deployment difficulties that prevent their successful implementation in our testing environment.

This selection enables us to compare our results with existing research and evaluate \app's effectiveness across a range of application domains.

\begin{table}[t]
\caption{REST Systems Used in the Evaluation.}
\label{table:services}
\begin{tabular}{lrr}
\hline
REST System & Lines of Code & \#Operations \\
\hline
Features Service & 1,646 & 18 \\
Rest Countries & 32,494 & 22 \\
News Service & 590 & 7 \\
SCS & 634 & 11 \\
NCS & 569 & 6 \\
Language Tool & 677,521 & 2 \\
Person Controller & 1,386 & 12 \\
Project Track & 88,634 & 67 \\
User Management & 5,108 & 22 \\
Genome Nexus & 22143 & 23 \\
Person Controller & 1,386 & 12 \\
PetStore & Online & 20 \\
Bill Service & Online & 21 \\
Genome Nexus  & Online & 23 \\
\hline
\end{tabular}
\vspace{-2mm}
\end{table}

\subsubsection{Baselines.}
We select four state-of-the-art REST API testing tools as baselines to study the effectiveness of \app in detecting server crashes (\ie \scode{500} errors) and test coverage:  
\begin{itemize}[leftmargin=12pt]
    \item RESTler~\cite{RESTler}: A stateful REST API testing tool open sourced by Microsoft.  
    \item EvoMaster~\cite{EvoMaster}: A search-based Web API testing tool based on evolutionary algorithms. We use its black-box mode in our experiments.
    \item Morest~\cite{Morest}: A model-based testing tool that utilizes a dynamic property graph to guide API call sequence generation.
    \item ARAT-RL~\cite{ARAT-RL}: A reinforcement learning-based approach that adaptively prioritizes operations and parameters to optimize the exploration process in REST API testing.  
\end{itemize}  

Since these baselines tools were originally designed for long-running testing, we conduct two experimental settings for each: one with a fixed number of API requests (1,000 calls) and another with a fixed time budget (1 hour), ensuring a fair and comprehensive performance comparison.

\subsubsection{Implementation.} For our experiments, we use GPT-4o-mini with default hyperparameters as the foundational LLM for all agents in \app, striking an optimal balance between model capabilities and computational cost. Before starting the workflow for each tested system, we restart the agents, clear the Scheduler and Memory, and reset the databases to ensure consistent initial states and eliminate cross-contamination between runs. Each system is tested by \app with a standardized budget of 1,000 API requests to ensure consistent evaluation across all test subjects. The experiments are conducted on cloud servers equipped with 8-core, 16-thread Intel Xeon Platinum processors, 32GB of RAM, and running Ubuntu 22.04.5 LTS.

\subsection{RQ1: Detection Effectiveness for Logical Issues}

\begin{table*}[htbp]
    \centering
    \caption{Effectiveness of \app in Detecting Logical Issues}
        \begin{tabular}{l|c|cc|cc|cc|cc}
            \toprule
            \multirow{2}{*}{Services} & \multicolumn{1}{c|}{Total Reports} & \multicolumn{2}{c|}{Bugs} & \multicolumn{2}{c|}{Enhancements} & \multicolumn{2}{c|}{All Issues} & \multicolumn{2}{c}{False-Positives} \\
            \cmidrule(lr){2-2} \cmidrule(lr){3-4} \cmidrule(lr){5-6} \cmidrule(lr){7-8} \cmidrule(lr){9-10}
             & Count & Count & Rate & Count & Rate & Count & Rate & Count & Rate \\
            \midrule
            Petstore & 41 & 20 & 48.78\% & 11 & 26.83\% & 31 & 75.61\% & 10 & 24.39\% \\
            Bill-Service & 23 & 7 & 30.43\% & 4 & 17.39\% & 11 & 47.83\% & 12 & 52.17\% \\
            Genome-Nexus & 49 & 27 & 55.10\% & 3 & 6.12\% & 30 & 61.22\% & 19 & 38.78\% \\
            Features-Service & 21 & 7 & 33.33\% & 8 & 38.10\% & 15 & 71.43\% & 6 & 28.57\% \\
            RESTCountries & 43 & 24 & 55.81\% & 16 & 37.21\% & 40 & 93.02\% & 3 & 6.98\% \\
            News-Service & 12 & 3 & 25.00\% & 5 & 41.67\% & 8 & 66.67\% & 4 & 33.33\% \\
            SCS & 22 & 15 & 68.18\% & 5 & 22.73\% & 20 & 90.91\% & 2 & 9.09\% \\
            NCS & 20 & 7 & 35.00\% & 7 & 35.00\% & 14 & 70.00\% & 6 & 30.00\% \\
            LanguageTool & 23 & 12 & 52.17\% & 3 & 13.04\% & 15 & 65.22\% & 8 & 34.78\% \\
            Person-Controller & 9 & 2 & 22.22\% & 2 & 22.22\% & 4 & 44.44\% & 5 & 55.56\% \\
            Project-Track & 48 & 7 & 14.58\% & 17 & 35.42\% & 24 & 50.00\% & 24 & 50.00\% \\
            User-Management & 38 & 8 & 21.05\% & 14 & 36.84\% & 22 & 57.89\% & 16 & 42.11\% \\
            \midrule
            \textbf{Total Count / Average\%} & 349 & 139 & 38.47\% & 95 & 27.71\% & 234 & 66.19\% & 115 & 33.81\% \\
            \bottomrule
        \end{tabular}
    \label{table:logic_issues}
    \vspace{-2mm}
\end{table*}

\subsubsection{Method}
For the potential logical issues reported by \app, manual verification is necessary, as they often rely on diverse oracles (\eg format requirements) and cannot be validated using status codes alone, unlike server crashes. 

For each logical issue report, we first collect all relevant information, including the request data, response data, the generated oracle, and \app's reasoning for why the response fails to meet expectations. The manual verification process begins with an analysis of the complete test scenario context—specifically, the sequence of all requests and responses within the given test case. This contextual examination is crucial for distinguishing true issues from false positives, particularly when incorrect oracles result from LLM hallucinations. Human annotators then conduct an in-depth review of \app's rationale to determine whether the reported issue is valid or a \textit{false positive}. Note that during manual verification, we also eliminate duplicate reports that share the same API and issue type.

If the logical issue is confirmed as valid, we classify its severity into two categories: \textbf{Enhancements} and \textbf{Bugs}. Enhancements refer to minor deviations that do not critically affect functionality, such as non-standard HTTP response codes or response structures differing from OpenAPI specifications, which may stem from inconsistencies in the specifications themselves. In contrast, bugs indicate severe logical inconsistencies, such as stored data becoming inaccessible or system behavior significantly deviating from expected functionality.

\subsubsection{Overall Results}

Table \ref{table:logic_issues} provides detailed results of \app's effectiveness in detecting logical issues. In total, \app reports 349 potential logical issues, of which \textbf{234 (66.19\%)} are confirmed as true system problems through rigorous manual verification. Specifically, \textbf{139 (38.47\%)} are classified as logical \textbf{bugs}, while \textbf{95 (27.71\%)} are identified as \textbf{enhancements} requiring system improvements. The remaining \textbf{115 (33.81\%)} are determined to be false positives. These results demonstrate \app's effectiveness, achieving a substantial detection accuracy of \textbf{66.19\%} across diverse REST systems while maintaining a reasonable false positive rate.

\subsubsection{Breakdown Analysis}
Beyond the overall results, a more granular analysis uncovers key patterns based on different characteristics of the tested systems.

\textit{\app exhibits higher false-positive rates in complex, domain-specific systems with numerous APIs and intricate business logic.}  
Notably, BillService (47.83\% accuracy, 12 false positives) and Project-Track (50.00\% accuracy, 24 false positives) illustrate this trend. Project-Track, with its 67 APIs, and BillService, which provides APIs for accessing and searching UK parliamentary bills and their legislative processes, highlight two key limitations of LLMs in handling domain-specific systems.
First, maintaining a comprehensive understanding is challenging when dealing with multiple interconnected APIs and complex specifications. Second, limited domain-specific knowledge leads to misinterpretations of intricate business rules, contributing to the higher false-positive rates in these complex systems.

\textit{\app can achieve high detection accuracy in certain poorly documented systems.}
For example, RESTCountries (93.02\% accuracy, 40 true issues), SCS (90.91\% accuracy, 20 true issues), and NCS (70\% accuracy, 14 true issues) exhibit high accuracy rates despite lacking details in their OpenAPI documentation, such as parameter specifications and response formats. 
This is due to the fact that LLMs understand a large amount of generalized knowledge, which allows it to reason through parts of the documentation to obtain the functionality that the system should have. At the same time, combined with the feedback from the testing process, it can reason about the undocumented information. Therefore, \app can achieve a high accuracy even in the absence of documentation.



\textit{A particularly interesting case: \app identifies 15 true issues in LanguageTool, which has only 2 APIs.}  
With only two APIs, LanguageTool focuses on language detection and grammar checking without complex business scenarios. However, its internal processing logic is highly intricate (677K LOC) due to the complexities of natural language analysis, particularly its multi-linguistic support. The relatively high number of detected issues likely stems from the LLM's strong linguistic knowledge and language processing capabilities, enabling \app to generate diverse test inputs (\eg multi-linguistic texts) and reliable oracles to assess LanguageTool's correctness and consistency. This is further supported by the significantly higher code coverage \app achieves for LanguageTool compared to other baselines (see Figure~\ref{fig:coverage_comparison}).


\subsection{RQ2: Detection Effectiveness for Server Crashes}

\subsubsection{Method}
Server crashes are a critical type of unhandled failure in REST systems and serve as the primary detection objective of baseline methods. To assess \app's effectiveness in identifying such failures, we systematically collect and analyze API responses that result in \scode{500} status codes across \app and baseline approaches. Specifically, for each API, we extract all responses that trigger a \scode{500} error and then deduplicate them based on response content.

\subsubsection{Results}
Table \ref{table:server_crashes} compares server crash detection results of different approaches. \app achieves the highest effectiveness within the same request budget (1,000 requests). When baselines are allowed to send significantly more requests within a fixed time duration, most—except Morest—achieve comparable results to \app by generating a large volume of requests. For instance, RESTler typically sends over 100K requests per hour for locally deployed systems, resulting in high system resource overhead and low exploration efficiency.

Notably, for Morest~\cite{Morest}, \scode{500} responses are deduplicated only based on APIs rather than response content, as its output does not include detailed response data.


\begin{table}[t]
    \setlength{\tabcolsep}{2.5pt}
    \caption{Distinct Server Crashes Detected by Different Tools}
    \resizebox{\columnwidth}{!}{
        \begin{tabular}{l|c|cc|cc|cc|cc}
            \toprule
            \multirow{2}{*}{Services} & \app & \multicolumn{2}{c|}{ARAT-RL} & \multicolumn{2}{c|}{EvoMaster} & \multicolumn{2}{c|}{Morest*} & \multicolumn{2}{c}{RESTler} \\
            \cmidrule(lr){2-2} \cmidrule(lr){3-4} \cmidrule(lr){5-6} \cmidrule(lr){7-8} \cmidrule(lr){9-10}
             & 1000 & 1000 & 1h & 1000 & 1h & 1000 & 1h & 1000 & 1h \\
            \midrule
            Petstore & 3 & 0 & 0 & 2 & 2 & 1 & 1 & 1 & 7 \\ 
            Bill-Service & 1 & 0 & 0 & 2 & 2 & 1 & 1 & 0 & 0 \\ 
            Genome-Nexus  & 0 & 0 & 0 & 2 & 3 & 0 & 0 & 0 & 0 \\ 
            Features-Service & 22 & 14 & 17 & 13 & 13 & 14 & 11 & 5 & 16 \\ 
            RESTCountries & 0 & 1 & 1 & 1 & 1 & 1 & 1 & 0 & 1 \\ 
            News-Service & 2 & 0 & 0 & 0 & 0 & 1 & 1 & 0 & 0 \\ 
            SCS & 0 & 0 & 0 & 0 & 0 & 0 & 0 & 0 & 0 \\ 
            NCS & 0 & 0 & 0 & 0 & 0 & 0 & 0 & 0 & 0 \\ 
            LanguageTool & 1 & 0 & 0 & 0 & 0 & 0 & 0 & 0 & 0 \\ 
            Person-Controller & 9 & 8 & 7 & 8 & 8 & 2 & 2 & 3 & 18 \\ 
            Project-Track & 3 & 4 & 15 & 5 & 7 & 2 & 3 & 3 & 5 \\ 
            User-Management & 8 & 5 & 5 & 6 & 7 & 6 & 6 & 6 & 7 \\ 
            \midrule
            Total & 49 & 32 & 45 & 39 & 43 & 28 & 26 & 18 & 54 \\ 
            \bottomrule
        \end{tabular}
    }
    \label{table:server_crashes}
    \vspace{-5mm}
\end{table}

\subsection{RQ3: Test Coverage}

\subsubsection{Method}
In this section, we compare the test coverage achieved by \app and the baselines, following prior studies~\cite{restgo,tosem_comparsion,coverage-1,coverage-2}. Test coverage is categorized into operation coverage and code coverage.  

Operation coverage applies to all tested systems. By analyzing the requests and responses generated during execution, we count successfully executed API calls (\ie those returning \scode{2xx} status codes) to assess operation coverage. This metric reflects how well a tool explores different APIs, providing a measure of its effectiveness in interacting with the tested system.

Code coverage applies only to REST systems that can be executed locally, as it requires inspecting their runtime behavior. We use JaCoCo~\cite{jacoco} to collect coverage data during service execution, measuring branch, line, and method-level coverage for both \app and the baselines. Code coverage quantifies how thoroughly a tool exercises the service’s internal logic, offering insights into its ability to explore different execution paths.

\subsubsection{Results}

\begin{figure}
    \centering
    \includegraphics[width=\columnwidth]{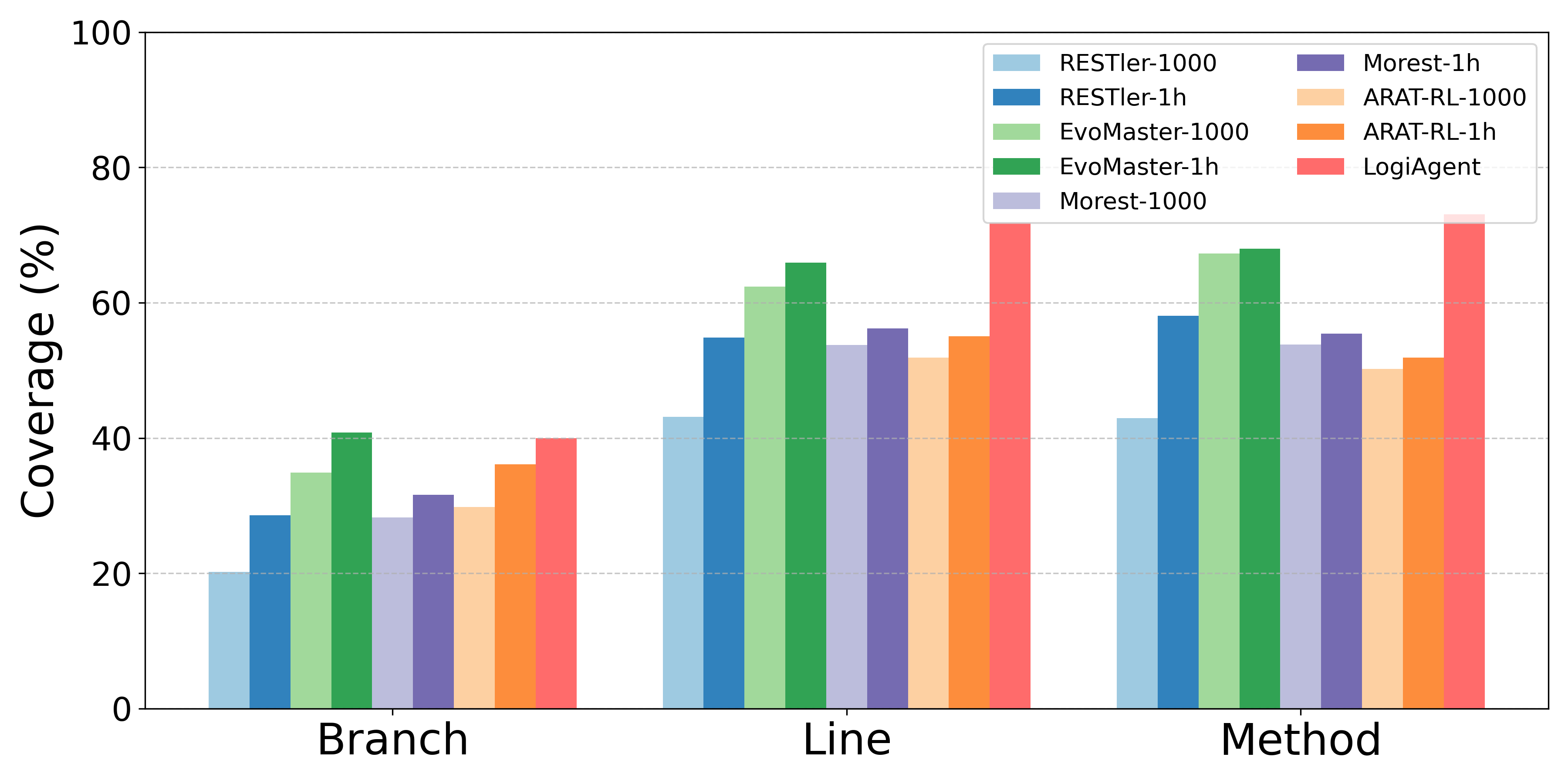}
    \caption{Average Code Coverage by Different Tools}
    \label{fig:coverage_average}
    \vspace{-2mm}
\end{figure}

The overall operation coverage and code coverage are presented in Table~\ref{table:operation_coverage} and Figure \ref{fig:coverage_average}, demonstrating \app's superior performance in achieving test coverage. 

\begin{table}[t]
    \setlength{\tabcolsep}{2.5pt}
    \caption{Operation Coverage by Different Tools}
    \resizebox{\columnwidth}{!}{
        \begin{tabular}{l|c|cc|cc|cc|cc}
            \toprule
            \multirow{2}{*}{Services} & \app & \multicolumn{2}{c|}{ARAT-RL} & \multicolumn{2}{c|}{EvoMaster} & \multicolumn{2}{c|}{Morest} & \multicolumn{2}{c}{RESTler} \\
            \cmidrule(lr){2-2} \cmidrule(lr){3-4} \cmidrule(lr){5-6} \cmidrule(lr){7-8} \cmidrule(lr){9-10}
             & 1000 & 1000 & 1h & 1000 & 1h & 1000 & 1h & 1000 & 1h \\
            \midrule
            Petstore & 17 & 15 & 15 & 14 & 16 & 13 & 17 & 1 & 18 \\ 
            Bill-Service & 17 & 3 & 7 & 2 & 12 & 4 & 14 & 0 & 1 \\ 
            Genome-Nexus  & 23 & 21 & 22 & 20 & 23 & 11 & 23 & 2 & 13 \\ 
            Features-Service & 18 & 10 & 17 & 7 & 7 & 9 & 14 & 4 & 16 \\ 
            RESTCountries & 22 & 6 & 14 & 17 & 22 & 8 & 18 & 2 & 22 \\ 
            News-Service & 7 & 2 & 3 & 2 & 3 & 2 & 2 & 2 & 7 \\ 
            SCS & 11 & 11 & 11 & 11 & 11 & 11 & 11 & 4 & 11 \\ 
            NCS & 6 & 6 & 6 & 6 & 6 & 6 & 6 & 5 & 6 \\ 
            LanguageTool & 2 & 1 & 2 & 1 & 1 & 1 & 1 & 2 & 2 \\ 
            Person-Controller & 8 & 4 & 5 & 12 & 12 & 7 & 7 & 1 & 12 \\ 
            Project-Track & 34 & 25 & 33 & 27 & 46 & 50 & 52 & 1 & 47 \\ 
            User-Management & 14 & 8 & 17 & 12 & 12 & 13 & 16 & 1 & 13 \\ 
            \midrule
            Total: 231 & 179 & 112 & 152 & 131 & 171 & 135 & 181 & 25 & 168 \\ 
            \bottomrule
        \end{tabular}
    }
    \label{table:operation_coverage}
\end{table}

\begin{figure}
    \centering
    \includegraphics[width=\columnwidth]{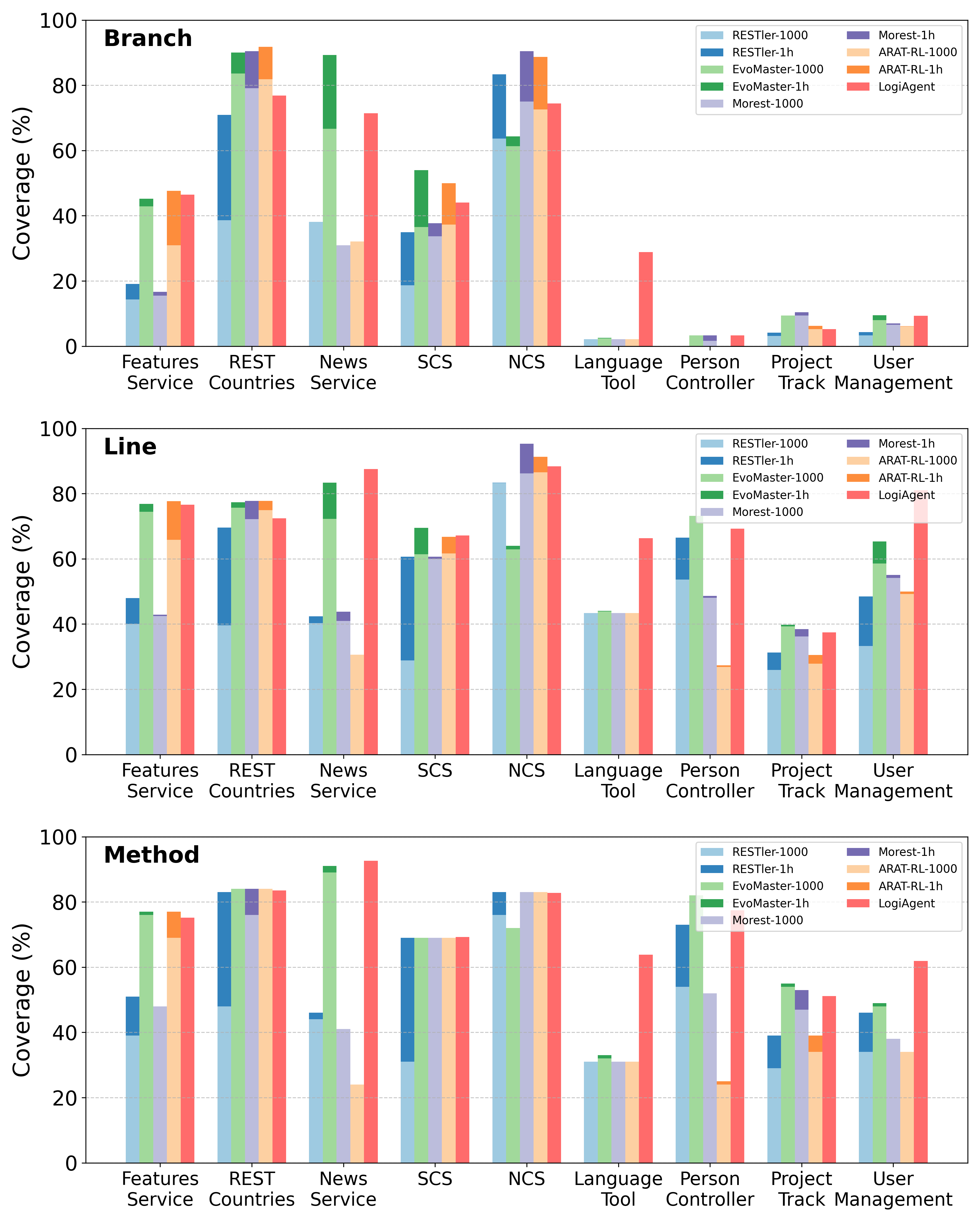}
    \caption{Code Coverage Comparison in Terms of Branch, Line, and Method}
    \label{fig:coverage_comparison}
    \vspace{-2mm}
\end{figure}

Figure~\ref{fig:coverage_comparison} presents system-specific code coverage. Compared to baseline tools limited to 1,000 requests (ARAT-RL-1000, Morest-1000, EvoMaster-1000, and RESTler-1000), \app consistently achieves higher coverage across all metrics. On average, \app attains 39.98\% branch coverage, 71.78\% line coverage, and 73.06\% method coverage, surpassing the best 1,000-request baseline results of 34.90\%, 62.38\%, and 67.24\%, respectively. Moreover, \app remains competitive even against tools running for one hour (denoted with “-1h”), which typically send tens to hundreds of thousands of requests. Despite using significantly fewer requests, \app achieves comparable or better performance. For example, its branch coverage (39.98\%) closely matches the highest one-hour result (40.83\%), while its line (71.78\%) and method (73.06\%) coverage exceed most one-hour baselines, with only EvoMaster-1h achieving similar results (65.92\% and 67.96\%).  

As noted in RQ1, LanguageTool is a particularly interesting case where \app achieves substantially higher code coverage than other baselines. This advantage is likely due to LLMs' strong language modeling and processing capabilities, enabling \app to generate diverse linguistic inputs and effective oracles for testing LanguageTool's behavior.

\subsection{RQ4: Ablation Study}

\subsubsection{Method}  
The ablation study evaluates the impact of the Execution Memory component on \app's code coverage. Specifically, we compare \app with two variants that omit key retrieval mechanisms:  
(i) \textit{w/o params}: The API Request Executor does not retrieve relevant parameters from Execution Memory.  
(ii) \textit{w/o refls}: The API Request Executor does not retrieve failure reflections from Execution Memory.

\subsubsection{Results}
Table~\ref{table:ablation} presents the ablation study results, showing that retrieval mechanisms in Execution Memory enhance code coverage by expanding the exploration space.

\begin{table}[t]
\centering
\caption{Results of the Ablation Study.}
\label{table:ablation}
\resizebox{\columnwidth}{!}{
\begin{tabular}{l|ccc}
\hline
Approaches & Branch & Line & Method \\
\hline
\app & 39.96\% & 71.75\% & 73.04\% \\
\quad - \textit{w/o params} & 35.80\% (-10.42\%) & 70.76\% (-1.38\%) & 71.67\% (-1.87\%) \\
\quad - \textit{w/o refls} & 37.72\% (-5.60\%) & 71.03\% (-1.00\%) & 71.84\% (-1.64\%) \\
\hline
\end{tabular}
}
\vspace{-2mm}
\end{table}

\section{Discussion}
\subsection{Limitations}
As an initial step towards logical testing for REST systems, \app shows promising potential, but its detection accuracy needs further improvement. False positives often result from LLM hallucinations when interpreting domain knowledge, understanding API specifications, and generating oracles. While LLMs can leverage general domain knowledge and common practices learned from public corpora, they still struggle with specific domains, especially those involving complex business logic. This is evident in the case of LanguageTool, where \app performs effectively despite the system's 677K LOC and only 2 APIs, as language analysis and processing are core capabilities of LLMs. However, for other domain-specific systems (\eg Bill-Service), \app exhibits higher false-positive rates. Future improvements could involve enhancing \app by constructing domain-specific knowledge bases, integrating not only API documentation but also the system's code implementation (if available).

\textit{}

\subsection{Threats to Validity}
Internal validity is challenged by human annotation of the reported logical issues. To mitigate this, we involve two experts with over 5 years of development experience in REST systems. The experts collaboratively resolve any conflicts in their annotations to reach consensus. Another internal validity concern is the flaky nature of REST systems, influenced by network stability. To address this, we deploy most of the tested systems locally and include three online systems to ensure a diverse range of scenarios are covered.

External validity primarily concerns the generalizability of our findings to a broader range of REST systems. To address this, we use widely-tested real-world systems across various application domains and include online systems to ensure that the testing reflects practical, real-world scenarios.
\section{Related Work}

In recent years, REST API has been widely used to build modern software systems. 
To guarantee the reliability of these systems, automated REST API testing has received extensive attention from researchers~\cite{restgo,tosem_survey,tosem_comparsion}.

RESTler~\cite{RESTler} and EvoMaster~\cite{EvoMaster} constitute early contributions in automated REST API testing and have been used in some industrial systems.
RESTler~\cite{RESTler} performs stateful testing of REST APIs by analyzing the producer-consumer relationship between request types.
EvoMaster~\cite{EvoMaster} is a search-based system testing tool that uses evolutionary algorithms to generate test cases.
It supports black-box testing and white-box testing of multiple Web APIs, such as REST, GraphQL, and RPC.
Several subsequent studies focus on mining and using operation or parameter constraints in REST APIs to guide the testing process.
Some studies uses graphs to model the dependencies between operations to guide the testing process~\cite{RestTestGen,Morest}.
RestTestGen~\cite{RestTestGen} identifies data dependencies between operations to construct an operation dependency graph, and craft call sequences based on the graph.
Morest~\cite{Morest} extends the operation dependency graph to a property graph and dynamically updates it.
Besides, some studies use parameter constraints to guide parameter generation~\cite{RESTest,RestCT,NLPtoREST}.
RESTest~\cite{RESTest} uses IDLReasoner to extract inter-parameter constraints for each API and uses the constraint solver to automatically generate valid test cases.
NLPtoREST~\cite{NLPtoREST} then uses natural language processing models to automatically identify inter-parameter constraints.
Some researches have adopted reinforcement learning in automated REST API testing.
ARAT-RL~\cite{ARAT-RL} uses Q-Learning to determine the priority of different operations and parameters during testing process.
DeepRest~\cite{DeepRest} combines deep reinforcement learning and curiosity-driven learning to explore the optimal call sequence exploration strategy and parameter generation strategy.

Recently, LLMs are widely used in software engineering tasks~\cite{llm4se-1,llm4se-2,llm4se-3}.
Some researchers have explored adopting LLMs in REST API testing.
RestGPT~\cite{RestGPT} enhances the REST API specifications by using LLMs to identify parameter constraints and generate parameter values.
AutoRestTest~\cite{AutoRestTest} uses few-shot learning to prompt LLMs to generate valid parameter values.
RESTSpecIT~\cite{RESTSpecIT} leverages LLMs to analyze the API response to automatically infer the API specification.
These explorations focus on leveraging LLMs for API parameter value generation and API response comprehension but do not examine their potential to enhance the overall testing process, particularly logical testing.

For efficient detection of system issues, test oracle is an important part of automated REST API testing.
Most of the existing work simply uses HTTP status codes as test oracle to validate the test result~\cite{RESTler,restler-data,EvoMaster,RestCT,Morest,DeepRest,ARAT-RL,RESTest}, which limits their ability in issue detection.
Dredd~\cite{Dredd} validates test result based on the schema of payloads by using the Gavel library~\cite{Gavel}.
AGORA~\cite{AGORA} uses Daikon to infer invariant properties from historical requests and responses, and uses them to validate the test result.
However, all of the existing works can only be used for validation of the test results of single REST API.
To the best of our knowledge, there is no existing work on testing and validation of logic issues in REST systems.

\section{Conclusion}
This paper introduces \app, a novel approach to logical testing for REST systems that overcomes the limitations of existing methods focused primarily on \scode{500} server errors. By utilizing an LLM-driven multi-agent framework, \app validates API responses with logical oracles grounded in business logic. Experiments across 12 real-world systems demonstrate \app's superior performance in detecting both logical issues and server crashes, outperforming state-of-the-art tools. Our future work will aim to enhance \app's capabilities by incorporating domain-specific knowledge for tested REST systems.

\clearpage
\bibliography{ref}
\bibliographystyle{ACM-Reference-Format}

\end{document}